\documentclass[english,aps,groupedaddress]{revtex4}
\usepackage{lmodern}

\usepackage[T1]{fontenc}
\usepackage[latin9]{inputenc}
\usepackage{babel}
\usepackage{array}
\usepackage{mathrsfs}
\usepackage{multirow}
\usepackage{amsmath}
\usepackage{amssymb}
\usepackage{graphicx}
\usepackage{esint}
\usepackage[unicode=true,pdfusetitle,
 bookmarks=true,bookmarksnumbered=false,bookmarksopen=false,
 breaklinks=false,pdfborder={0 0 1},backref=false,colorlinks=false]
 {hyperref}

\makeatletter

\providecommand{\tabularnewline}{\\}

\@ifundefined{textcolor}{}
{%
 \definecolor{BLACK}{gray}{0}
 \definecolor{WHITE}{gray}{1}
 \definecolor{RED}{rgb}{1,0,0}
 \definecolor{GREEN}{rgb}{0,1,0}
 \definecolor{BLUE}{rgb}{0,0,1}
 \definecolor{CYAN}{cmyk}{1,0,0,0}
 \definecolor{MAGENTA}{cmyk}{0,1,0,0}
 \definecolor{YELLOW}{cmyk}{0,0,1,0}
}

\usepackage[paperwidth=210mm,paperheight=297mm,centering,hmargin=2cm,vmargin=2.5cm]{geometry}

\makeatother

\begin{document}

\title{Cosmological Perturbations in Antigravity}

\author{Marius Oltean}

\email{moltean@physics.mcgill.ca}

\selectlanguage{english}%

\affiliation{Department of Physics, McGill University, Montreal, Quebec, H3A 2T8,
Canada}

\author{Robert Brandenberger}

\email{rhb@hep.physics.mcgill.ca}

\selectlanguage{english}%

\affiliation{Department of Physics, McGill University, Montreal, Quebec, H3A 2T8,
Canada}

\date{\today}
\begin{abstract}
We compute the evolution of cosmological perturbations in a recently
proposed Weyl-symmetric theory of two scalar fields with oppositely-signed
conformal couplings to Einstein gravity. It is motivated from the
minimal conformal extension of the Standard Model, such that one of
these scalar fields is the Higgs while the other is a new particle,
the dilaton, introduced to make the Higgs mass conformally symmetric.
At the background level, the theory admits novel geodesically-complete
cyclic cosmological solutions characterized by a brief period of repulsive
gravity, or \textquotedblleft{}antigravity,\textquotedblright{} during
each successive transition from a Big Crunch to a Big Bang. We show
that despite the necessarily wrong-signed kinetic term of the dilaton
in the full action, its cosmological solutions are stable at the perturbative
level.
\end{abstract}
\maketitle

\section{Introduction}

The Big Bang singularity is arguably the most critical problem at
the heart of modern cosmology. In the context of Einstein gravity
minimally coupled to ordinary matter --- the standard setting for
the preponderance of current early Universe scenarios, such as inflation
--- initial singularities are unavoidable. To wit, their ubiquity
has long been understood to be inextricably incurred by singularity
theorems \cite{Hawking1970} which prove cosmological spacetimes therein
to be, necessarily, geodesically-incomplete \cite{Hawking1973,Wald1984}.

Thus, in the present absence of a theory of quantum gravity, one potentially
fruitful approach to constructing geodesically-complete cosmologies
is the consideration of cosmological solutions to modified theories
of (classical) gravity. In this paper, we will focus our attention
on one particular such theory: namely, a scalar-tensor theory recently
proposed and developed on the basis of Weyl symmetry \cite{Bars2010,Bars2011a,Bars2011b,Bars2011c,Bars2012a,Bars2012b,Bars2013a,Bars2013b,Bars:2013c}
\footnote{See also \cite{Padilla} for a discussion of more general two
field scale-invariant theories coupled to gravity.}.
As its principal motivation, it has been argued to arise from the
simplest Weyl-symmetric coupling of Einstein gravity to the minimal
conformally-invariant extension of the (classical) Standard Model
\cite{Bars2013a}. This leads (modulo the remaining particle content
of the usual Standard Model) to an action for two scalars, the Higgs
$s$ and a new field introduced to make its mass conformally symmetric,
the dilaton $\phi$, each conformally coupled to Einstein gravity
but with opposite signs. 
This allows for solutions to transit
from a gravity phase to an antigravity phase and back 
\footnote{In the context of single scalar field matter, models
with non-minimal coupling between matter and gravity which allow
for solutions to transit between the gravity and antigravity
phase were studied in \cite{Linde}, but they were shown to
have singularities and instabilities \cite{Starob, Futamase, Raul}.}.

While reducing to known physics in the expected limits (to ordinary
Einstein gravity at low energies, and to the Standard Model in flat
space), this theory entails new effects in regimes with strong gravity
and large field fluctuations (when the amplitudes of $\phi$ and $s$
are comparable). Such conditions arise, quite naturally, in cosmology
--- specifically, in the vicinity of the Big Bang. Concordantly, a
wealth of novel cosmological solutions have been obtained in this
theory \cite{Bars2012a}; they are all geodesically-complete, and
they all repeat cyclically (through Big Crunch-Big Bang transitions)
\footnote{Note that some curvature invariants blow up at
the transition point between the gravity and antigravity phase
(see e.g. \cite{Kallosh}). However, it was argued in \cite{BSTnew}
that this does not effect the viability of the geodesically
complete solutions.}.

The qualitative features \cite{Bars2012b} of these (homogeneous)
solutions depend on whether or not anisotropies are included. For
ease of discussion, we introduce the following definition for dividing
the field space $\left\{ \left(\phi,s\right)\right\} $ into two regions:
\begin{equation}
\mathbb{G}_{\pm}=\left\{ \left(\phi,s\right):\left|\phi\right|\gtrless\left|s\right|\right\} .
\end{equation}
In other words, in $\mathbb{G}_{+}$, the amplitude of the Higgs is
\textsl{less} than that of the dilaton; by construction of their respective
conformal couplings, this corresponds to a \textsl{positive} Newton
constant, and therefore gravity being \textsl{attractive}. Meanwhile,
in $\mathbb{G}_{-}$, the amplitude of the Higgs is \textsl{greater}
than that of the dilaton; this leads, accordingly, to a \textsl{negative}
Newton constant, and therefore gravity being \textsl{repulsive} ---
also referred to as \textsl{antigravity}.

\begin{figure}[h]
\includegraphics[scale=0.3]{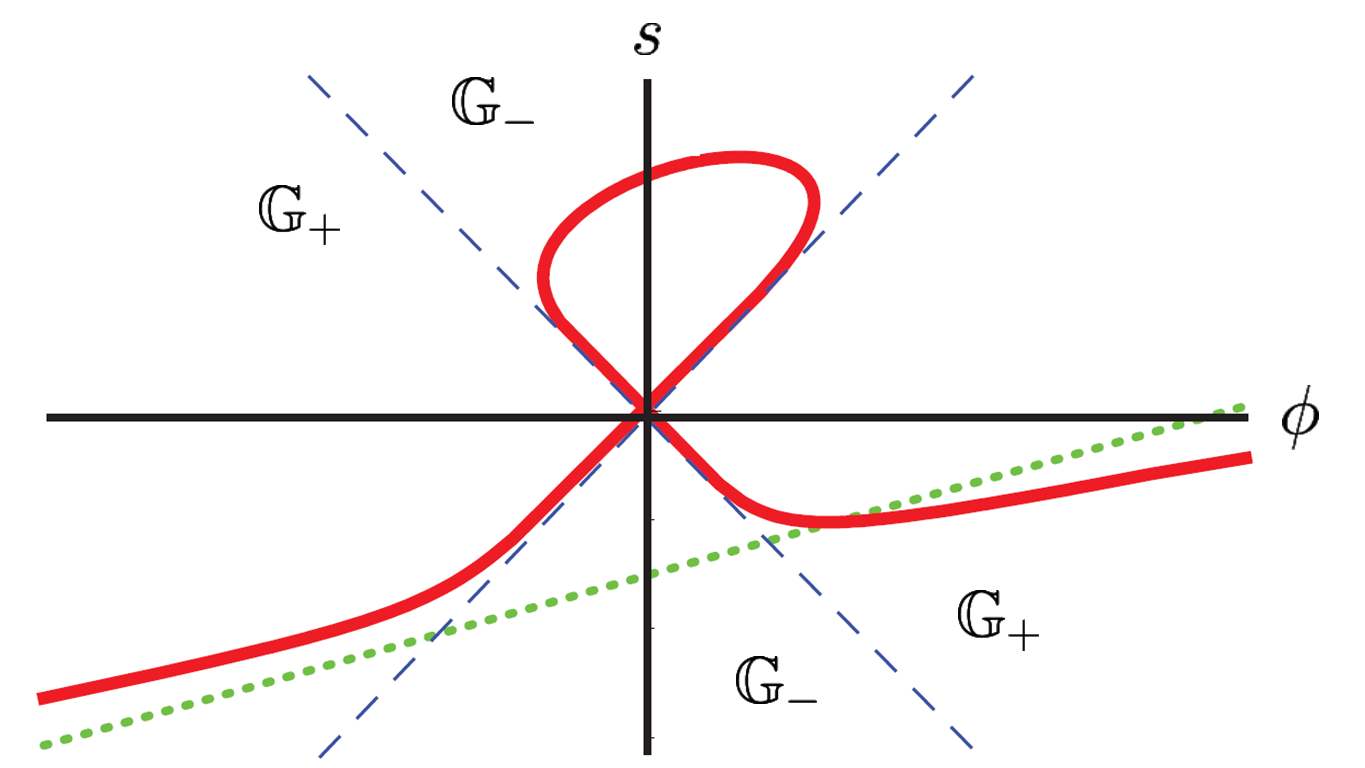}\caption{\label{fig:BackgroundSol}Taken from \cite{Bars2012a}. Generic background
solutions without anisotropy (dotted green) and with anisotropy (solid
red). The dashed blue lines bound the regions where gravity is attractive
($\mathbb{G}_{+}$) and repulsive ($\mathbb{G}_{-}$).}
\end{figure}

Without anisotropy, generic solutions (meaning, for varied choices
of initial conditions and parameters of the model) cyclically pass
through zero-size bounces: From a lengthy period in $\mathbb{G}_{+}$,
they cross (at the Big Crunch) into a brief period in $\mathbb{G}_{-}$,
and then exit it (at the following Big Bang) into another lengthy
period in $\mathbb{G}_{+}$, and so on. A typical solution of this
sort is the dotted green curve in Figure \ref{fig:BackgroundSol}.
(There is in fact also a special class of these solutions which can
go through zero-size bounces --- or even finite-size bounces, if they
are spatial curvature-induced --- without actually having to pass
through $\mathbb{G}_{-}$.)

With anisotropy, an attractor mechanism is created (independent of
initial conditions or choice of potential). All solutions are (cyclically)
forced, after a lengthy period in $\mathbb{G}_{+}$, through the origin
of field space (contracting to zero size at the Big Crunch), and then
into a loop in $\mathbb{G}_{-}$, finally exiting it again through
the origin (once more contracting to zero size) back into $\mathbb{G}_{+}$
(leading to the usual expanding phase of attractive gravity following
the Big Bang), and so on. A typical depiction of such a solution is
the solid red curve in Figure \ref{fig:BackgroundSol}.

Notwithstanding the merits of this theory insofar as it is able to
resolve geodesic completeness in cosmology, the necessarily opposite
signs of the conformal couplings of the two scalar fields constituting
it immediately raises the question of ghosts, and the extent to which
such may present any danger. In particular, although the Higgs kinetic
term has the correct sign in the action (as it ought to), the kinetic
term for the dilaton has the ``wrong'' sign, explicitly making it
--- by construction --- a ghost. A priori, this need not be construed
as problematic at least at the background level, since a unitary gauge
choice (permitted by the Weyl symmetry) can be used to eliminate the
ghost degree of freedom. It is not immediately obvious, however, if
perturbations in this theory are themselves ghost-free. In other words,
we cannot deduce right away whether the kinetic term of the second-order
action will have the correct sign (nor the mass term, for that matter).
If it does not, this \textsl{would }be problematic for the cosmological
solutions of this theory \cite{Cline2004}. Thus, the principal aim
of this paper will be to compute the evolution of its scalar perturbations
with the motivating aim of establishing their stability (i.e. whether
they are free of ghosts/tachyons). For simplicity, we will perform
the analysis here in the absence of anisotropies, with potential
set to zero and without any radiation. In addition, we do not
consider the gravitational wave sector.

The remainder of this paper is structured as follows. In Section \ref{sec:Setup}
we give the full action and equations of motion for this theory. Then,
in Section \ref{sec:Cosmological-Perturbations}, we develop our treatment
of its cosmological perturbations. Following this, we analyze their
stability in Section \ref{sec:Stability-Analysis}, and finally, in
Section \ref{sec:Conclusions}, we offer concluding remarks.

\section{Setup\label{sec:Setup}}

We work in the $\left(+,-,-,-\right)$ metric signature, and write
all dimensionful quantities in Planck units.

As noted, the action for this theory comprises two scalar fields,
the dilaton $\phi$ and the Higgs $s$, conformally coupled with opposite
signs to Einstein gravity:
\begin{equation}
S\left[\phi,s;g_{\mu\nu}\right]=\int{\rm d}^{4}x\sqrt{-g}\bigg\{-\frac{\left(\phi^{2}-s^{2}\right)}{12}R-\frac{1}{2}g^{\mu\nu}\Big(\nabla_{\mu}\phi\nabla_{\nu}\phi-\nabla_{\mu}s\nabla_{\nu}s\Big)\bigg\}.\label{eq:S}
\end{equation}
A potential $V\left(\phi,s\right)$ for the two scalars may also be
added to this, but for simplicity we will work with it set to zero.

The gravitational equation of motion is obtained by varying (\ref{eq:S})
with respect to the metric, which yields:
\begin{equation}
\left(\phi^{2}-s^{2}\right)G_{\alpha\beta}+g_{\alpha\beta}g^{\mu\nu}\nabla_{\mu}\nabla_{\nu}\left(\phi^{2}-s^{2}\right)-\nabla_{\alpha}\nabla_{\beta}\left(\phi^{2}-s^{2}\right)=-6T_{\alpha\beta},\label{eq:Gravity}
\end{equation}
where on the right-hand side we have the usual stress-energy tensor
of the two scalar fields (with appropriate opposite signs),
\begin{equation}
T_{\alpha\beta}=-\frac{1}{2}g_{\alpha\beta}g^{\mu\nu}\Big(\nabla_{\mu}\phi\nabla_{\nu}\phi-\nabla_{\mu}s\nabla_{\nu}s\Big)+\Big(\nabla_{\alpha}\phi\nabla_{\beta}\phi-\nabla_{\alpha}s\nabla_{\beta}s\Big).
\end{equation}

Meanwhile, varying (\ref{eq:S}) with respect to either of the two
scalars produces the same matter equation of motion,
\begin{equation}
\phi g^{\mu\nu}\nabla_{\mu}\nabla_{\nu}s-sg^{\mu\nu}\nabla_{\mu}\nabla_{\nu}\phi=0.\label{eq:Matter}
\end{equation}
The fact that there is not a separate equation of motion for each
scalar field happens to be a consequence of the Weyl symmetry along
with the fact that we have chosen a vanishing potential \cite{Bars2010}.

\section{Cosmological Perturbations\label{sec:Cosmological-Perturbations}}

We work in the conformal Newtonian gauge, and assume that
there is no anistropic stress
at linear order in the matter fluctuations.
This allows us to employ the perturbed Friedmann-Robertson-Walker
metric
\begin{equation}
g_{\mu\nu}=a^{2}\left(\tau\right)\left[\eta_{\mu\nu}+2\varepsilon\Phi\left(\tau,\mathbf{x}\right)\delta_{\mu\nu}\right],\label{eq:g}
\end{equation}
where $\eta_{\mu\nu}={\rm diag}\left(1,-1,-1,-1\right)$ is the Minkowski
metric, $\Phi$ is the scalar metric perturbation, and we use $\varepsilon$
as our perturbation parameter (so that $\mathcal{O}(\varepsilon^{n})$
is the $n$-th perturbative order). Moreover, we consider perturbed
expressions for the two matter fields,
\begin{equation}
\begin{cases}
\phi\left(\tau,\mathbf{x}\right) & =\phi_{0}\left(\tau\right)+\varepsilon\Pi\left(\tau,\mathbf{x}\right),\\
s\left(\tau,\mathbf{x}\right) & =s_{0}\left(\tau\right)+\varepsilon\Theta\left(\tau,\mathbf{x}\right),
\end{cases}\label{eq:phi&s}
\end{equation}
where $\phi_{0}$ and $s_{0}$ are the background (i.e. homogeneous)
solutions for the dilaton and Higgs fields, while $\Pi$ and $\Theta$
are their perturbations, respectively.

Furthermore, it will be useful for later convenience to define the
following background variables:
\begin{equation}
\begin{cases}
z & =\phi_{0}^{2}-s_{0}^{2},\\
\tilde{z} & =\phi_{0}'^{2}-s_{0}'^{2},
\end{cases}\label{eq:z}
\end{equation}
as well as perturbation variables:
\begin{equation}
{\displaystyle \begin{cases}
\alpha & =\phi_{0}\Pi-s_{0}\Theta,\\
\tilde{\alpha} & =\phi_{0}'\Pi'-s_{0}'\Theta',
\end{cases}}\quad{\rm and}\quad{\displaystyle \begin{cases}
\Upsilon & =z\Phi-\alpha,\\
\tilde{\Upsilon} & =\tilde{z}\Phi-\tilde{\alpha}.
\end{cases}}\label{eq:a&U}
\end{equation}
We can now proceed to studying the evolution of these perturbations.
We approach this both from the point of view of the equations of motion,
and of the action. As a consistency check, we show in Appendix \ref{a:Consistency}
that all of our results detailed in the following subsections correctly
reduce to those of the standard theory of cosmological perturbations
in Einstein gravity in the appropriate limit.

\subsection{Perturbed Equations of Motion}

We derive the perturbed gravitational and matter equations of motion
by directly inserting (\ref{eq:g}) and (\ref{eq:phi&s}) into (\ref{eq:Gravity})
and (\ref{eq:Matter}). We state the results in Table \ref{tab:1},
making use of the variables (\ref{eq:z}) and (\ref{eq:a&U}) defined
above. In the case of Einstein gravity, the off-diagonal space-space 
equations are trivial in the absence of anisotropic stress. This is
not the case here. The off-diagonal space-space equation of motion
reads
\begin{equation}
\phi_{0}\partial_{i}\partial_{j}\Pi 
- s_{0}\partial_{i}\partial_{j}\Theta = 0 \, .
\end{equation}
We shall assume that this constraint equation is satisfied. This
assumption is only possible to make because we have two matter
fields. In the case of a single non-canonically coupled matter field
this equation cannot be satisfied and leads to a paradox \cite{Caputa}.

\begin{table}[h]
\begin{tabular}{|>{\raggedright}m{2.5cm}||>{\raggedright}m{3.5cm}|>{\raggedright}m{6.5cm}|}
\hline 
\multirow{2}{2.5cm}{EOM} & \multirow{2}{3.5cm}{~$\mathcal{O}\left(1\right)$} & \multirow{2}{6.5cm}{~$\mathcal{O}\left(\varepsilon\right)$}\tabularnewline
 &  & \tabularnewline
\hline 
\hline 
\multirow{2}{2.5cm}{(\ref{eq:Gravity}) time-time} & \multirow{2}{3.5cm}{~$0=z\mathcal{H}^{2}+z'\mathcal{H}+\tilde{z}$} & \multirow{2}{6.5cm}{~$0=\nabla^{2}\Upsilon-3\mathcal{H}\left(\Upsilon'+\Upsilon\mathcal{H}\right)-3\tilde{\Upsilon}-\frac{3}{2}z'\Phi'$}\tabularnewline
 &  & \tabularnewline
\hline 
\multirow{4}{2.5cm}{(\ref{eq:Gravity}) space-space} & \multirow{4}{3.5cm}{~$0=z\left(2\mathcal{H}'+\mathcal{H}^{2}\right)$\\
\vspace{0.05in}
$\quad\quad+z'\mathcal{H}+z''-3\tilde{z}$} & \multirow{4}{6.5cm}{~$0=\Upsilon''+\mathcal{H}\left(\Upsilon'+2z\Phi'\right)+\frac{1}{2}z'\Phi'$\\
\vspace{0.05in}
$\quad\quad+\left(2\mathcal{H}'+\mathcal{H}^{2}\right)\Upsilon-3\tilde{\Upsilon}+\frac{2}{3}\nabla^{2}\alpha$}\tabularnewline
 &  & \tabularnewline
 &  & \tabularnewline
 &  & \tabularnewline
\hline 
\multirow{4}{2.5cm}{(\ref{eq:Gravity}) time-space} & \multirow{4}{3.5cm}{~Trivial.} & \multirow{4}{6.5cm}{~$0=\partial_{i}\big[z\left(\mathcal{H}\Phi+\Phi'\right)+\frac{1}{2}z'\Phi+\mathcal{H}\alpha-\alpha'$\\
\vspace{0.05in}
$\quad\quad\quad+3(\phi_{0}'\Pi-s_{0}'\Theta)\big]$}\tabularnewline
 &  & \tabularnewline
 &  & \tabularnewline
 &  & \tabularnewline
\hline 
\multirow{5}{2.5cm}{(\ref{eq:Matter})} & \multirow{5}{3.5cm}{~$0=\phi_{0}\left(s_{0}''+2\mathcal{H}s_{0}'\right)$\\
\vspace{0.05in}
$\quad\quad-s_{0}\left(\phi_{0}''+2\mathcal{H}\phi_{0}'\right)$} & \multirow{5}{6.5cm}{~$0=\phi_{0}\left(\Theta''-\nabla^{2}\Theta+2\mathcal{H}\Theta'-4s_{0}'\Phi'\right)$\\
\vspace{0.05in}
$\quad\quad-s_{0}\left(\Pi''-\nabla^{2}\Pi+2\mathcal{H}\Pi'-4\phi_{0}'\Phi'\right)$\\
\vspace{0.05in}
$\quad\quad-\left(\phi_{0}''+2\mathcal{H}\phi_{0}'\right)\Theta+\left(s_{0}''+2\mathcal{H}s_{0}'\right)\Pi$}\tabularnewline
 &  & \tabularnewline
 &  & \tabularnewline
 &  & \tabularnewline
 &  & \tabularnewline
\hline 
\end{tabular}\caption{\label{tab:1}Perturbed equations of motion.}
\end{table}

While direct computation of these equations offers one possible approach
to studying the evolution of perturbations in this theory, they are
not particularly illuminating especially when it comes to the question
of how we might reduce the number of propagating degrees of freedom
at the perturbative level from the current three (the metric perturbation
$\Phi$, the dilaton perturbation $\Pi$ and the Higgs perturbation
$\Theta$). More helpful in this regard will be to compute the second-order
action.

\subsection{Perturbed Action}

Inserting (\ref{eq:g}) and (\ref{eq:phi&s}) into (\ref{eq:S}) and
simplifying the result as much as possible via integration by parts
and substitution of background equations (i.e. the $\mathcal{O}(1)$
equations from Table \ref{tab:1}), we obtain the following second-order
action:
\begin{align}
S^{\left(2\right)}=\varepsilon^{2}\int\mathrm{d}^{4}x\,\frac{a^{2}}{2}\bigg\{ & -z\left[\Phi'^{2}+\frac{1}{3}\left(\nabla\Phi\right)^{2}\right]-\left[\Pi'^{2}-\Theta'^{2}\right]+\left[\left(\nabla\Pi\right)^{2}-\left(\nabla\Theta\right)^{2}\right]+\left(\mathcal{H}^{2}+\mathcal{H}'\right)\left[\Pi^{2}-\Theta^{2}\right]\nonumber \\
 & -\frac{2}{3}\left(\nabla\alpha\cdot\nabla\Phi\right)+2\Phi'\left(\alpha'+2\mathcal{H}\alpha\right)+8\Phi\left[\mathcal{H}\left(\alpha'+\mathcal{H}\alpha\right)+\tilde{\alpha}\right]\bigg\}.\label{eq:S2}
\end{align}
As expected, we have three kinetic terms corresponding to each of
the three perturbation variables. We can now try to further simplify
this with the help of the time-space gravitational equation.

\subsubsection{From Two to One Matter Perturbation}

Consider the (\ref{eq:Gravity}) time-space $\mathcal{O}\left(\varepsilon\right)$
equation from Table \ref{tab:1}, which we rewrite more suggestively
as:
\begin{equation}
\phi_{0}\Pi'-s_{0}\Theta'=\frac{1}{3}\left[z\left(\mathcal{H}\Phi+\Phi'\right)+\frac{1}{2}z'\Phi+\mathcal{H}\alpha+2\alpha'\right].\label{eq:Constraint}
\end{equation}
This indicates that one out of the three propagating degrees of freedom
appearing in (\ref{eq:S2}) can in fact be expressed in terms of the
other two; in other words, it should be possible to get the perturbed
action in terms of only two perturbation variables. To do this, we
proceed as follows: We write $\Pi=(\alpha+s_{0}\Theta)/\phi_{0}$
(by the definition (\ref{eq:a&U}) of $\alpha$), take its time derivative,
and insert the resulting expression for $\Pi'$ into (\ref{eq:Constraint}).
The $\Theta'$ terms cancel, and we get an expression for $\Theta$,
\begin{equation}
\Theta=\left[\frac{z\mathcal{H}+\frac{1}{2}z'}{3\phi_{0}\left(s_{0}/\phi_{0}\right)'}\right]\Phi+\left[\frac{z}{3\phi_{0}\left(s_{0}/\phi_{0}\right)'}\right]\Phi'-\left[\frac{\alpha}{3\phi_{0}\left(s_{0}/\phi_{0}\right)'}\right]'+\left[3\left(\frac{\phi_{0}'}{\phi_{0}}+\mathcal{H}\right)+\frac{\phi_{0}'}{\phi_{0}}\right]\frac{\alpha}{3\phi_{0}\left(s_{0}/\phi_{0}\right)'},\label{eq:T}
\end{equation}
simply in terms of $\Phi$, $\alpha$, and their derivatives. Then,
$\Pi$ can in turn be written in terms of the same by putting this
expression for $\Theta$ into $\Pi=(\alpha+s_{0}\Theta)/\phi_{0}$.

To make these expressions for the matter perturbations cleaner, as
well as for future convenience, we define a list of additional background
variables in Table \ref{tab:2}.

\begin{table}[h]
\begin{tabular}{|>{\raggedright}m{4cm}||>{\raggedright}m{4cm}|}
\hline 
\multirow{3}{4cm}{~${\displaystyle u=\frac{s_{0}}{\phi_{0}}}$} & \multirow{3}{4cm}{~${\displaystyle v=\frac{\phi_{0}'}{\phi_{0}}}$}\tabularnewline
 & \tabularnewline
 & \tabularnewline
\hline 
\multirow{3}{4cm}{~${\displaystyle \tilde{u}=1-u^{2}=\frac{z}{\phi_{0}^{2}}}$} & \multirow{3}{4cm}{~${\displaystyle \tilde{v}=v+\mathcal{H}}$}\tabularnewline
 & \tabularnewline
 & \tabularnewline
\hline 
\multirow{3}{4cm}{~${\displaystyle \bar{u}=\frac{\phi_{0}\tilde{u}}{6u'}}$} & \multirow{3}{4cm}{~${\displaystyle \bar{v}=4\tilde{v}v-2\tilde{v}'+\mathcal{H}^{2}+\mathcal{H}'}$}\tabularnewline
 & \tabularnewline
 & \tabularnewline
\hline 
\multirow{3}{4cm}{~${\displaystyle \hat{u}=v\tilde{u}'-\left(\mathcal{H}^{2}+\mathcal{H}'\right)\tilde{u}}$} & \multirow{3}{4cm}{~${\displaystyle \hat{v}=6\left(\tilde{v}\tilde{u}'-u'^{2}\right)}$}\tabularnewline
 & \tabularnewline
 & \tabularnewline
\hline 
\end{tabular}\caption{\label{tab:2}Useful background variables.}

\end{table}

Now, (\ref{eq:T}) readily suggests that we work not with the variable
$\alpha=\phi_{0}\Pi-s_{0}\Theta$, but instead with
\begin{equation}
\psi=\frac{\alpha}{3\phi_{0}u'}=\frac{\phi_{0}\Pi-s_{0}\Theta}{3\phi_{0}\left(s_{0}/\phi_{0}\right)'}.\label{eq:psi}
\end{equation}
Combining all of the above, we can express both (dilaton and Higgs)
matter perturbations simply in terms of the metric perturbation $\Phi$,
the new variable $\psi$, and their derivatives (as well as background
variables):
\begin{equation}
\begin{cases}
\Pi & =u\Theta+3u'\psi,\\
\Theta & =2\bar{u}\Phi'+\left(\bar{u}'-v\bar{u}\right)\Phi-\psi'+\left(3\tilde{v}+v\right)\psi.
\end{cases}\label{eq:P&T}
\end{equation}
Henceforth, we can regard $\psi$ as being the (single) variable describing
matter perturbations.

\subsubsection{Perturbed Action in Two Variables}

The next step is to insert (\ref{eq:P&T}) into the second order action
(\ref{eq:S2}) and thereby write the latter purely in terms of $\Phi$
and $\psi$.

The first step in the simplification is to get rid of the $\tilde{\alpha}$
term. To do this, we insert its definition (\ref{eq:a&U}) in terms
of the original perturbation variables, $\tilde{\alpha}=\phi_{0}'\Pi'-s_{0}'\Theta'$,
integrate it by parts, and then use (\ref{eq:Constraint}) to simplify:
\begin{align}
S^{\left(2\right)}=\varepsilon^{2}\int\mathrm{d}^{4}x\,\frac{a^{2}}{2}\Big\{ & -\Pi'^{2}+\Theta'^{2}+\left[\left(\nabla\Pi\right)^{2}-\left(\nabla\Theta\right)^{2}\right]+\left[\left(\mathcal{H}^{2}+\mathcal{H}'\right)\left(\Pi^{2}-\Theta^{2}\right)\right]+z\Big[\frac{5}{3}\Phi'^{2}-\frac{1}{3}\left(\nabla\Phi\right)^{2}\Big]\nonumber \\
 & -\frac{2}{3}\left(\nabla\alpha\cdot\nabla\Phi\right)-\frac{2}{3}\Phi'\alpha'-\frac{4}{3}\mathcal{H}\Phi'\alpha-8\left[v'+\left(v+2\mathcal{H}\right)v+\mathcal{H}^{2}+\mathcal{H}'\right]\Phi\alpha\Big\}.
\end{align}
To make progress, we set at this point all spatial gradients to zero.
Then, writing $\alpha$ in terms of the more convenient variable $\psi$
defined by (\ref{eq:psi}), the above becomes:
\begin{align}
S^{\left(2\right)}=\varepsilon^{2}\int\mathrm{d}^{4}x\,\frac{a^{2}}{2}\Big\{ & -\Pi'^{2}+\Theta'^{2}+\left[\mathcal{H}^{2}+\mathcal{H}'\right]\left(\Pi^{2}-\Theta^{2}\right)+\frac{5}{3}z\Phi'^{2}\nonumber \\
 & -2\phi_{0}u'\Phi'\left(\psi'-v\psi\right)-24\phi_{0}u'\left[\tilde{v}'+\tilde{v}^{2}\right]\Phi\psi\Big\}.
\end{align}
Now, we can write $\Pi$ in terms of $\Theta$ using the first expression
in (\ref{eq:P&T}):
\begin{align}
S^{\left(2\right)}=\varepsilon^{2}\int\mathrm{d}^{4}x\,\frac{a^{2}}{2}\Big\{ & \tilde{u}\Theta'^{2}+\hat{u}\Theta^{2}-3\bar{v}\tilde{u}'\psi\Theta+\hat{v}\psi'\Theta+3\tilde{u}'\psi'\Theta'-9u'^{2}\psi'^{2}+9\bar{v}u'^{2}\psi^{2}\nonumber \\
 & +\frac{5}{3}z\Phi'^{2}-2\phi_{0}u'\Phi'\left(\psi'-v\psi\right)-24\phi_{0}u'\left[\tilde{v}'+\tilde{v}^{2}\right]\Phi\psi\Big\}.\label{eq:3.S2preABCD}
\end{align}
Finally, we could substitute into this our second expression from
(\ref{eq:P&T}) for $\Theta$ in terms of $\Phi$, $\psi$ and their
derivatives, in order to get $S^{\left(2\right)}$ entirely in terms
of the latter two variables. However, there is an issue with this:
Since $\Theta$ contains $\Phi'$ and $\psi'$ terms, and since $S^{\left(2\right)}$
contains $\Theta'^{2}$ terms, this will lead to $\Phi''^{2}$ and
$\psi''^{2}$ terms in the action. Instead, we would like to have
just standard kinetic-type (i.e. single time derivative squared) terms.
Accordingly, this difficulty immediately suggests a possible resolution.

\subsubsection{Field Redefinitions and Diagonalization}

If we were to define a new field as a linear combination of $\Phi$
and $\Phi'$, then the $\Phi''^{2}$ term appearing in $S^{\left(2\right)}$
might simply be construed as a contribution to its (standard) kinetic-type
term --- and similarly in the case of $\psi$. Alternately, we can
take $\Phi'$ to be a linear combination of the new field and $\Phi$
--- and again, similarly for $\psi$.

Following this logic, let $A$, $B$, $C$, and $D$ be (as of yet)
undetermined functions of time, and define two new fields $\tilde{\Phi}$
and $\tilde{\psi}$ according to $\Phi'=A\tilde{\Phi}+B\Phi$ and
$\psi'=C\tilde{\psi}+D\psi$, or:
\begin{equation}
\begin{cases}
\tilde{\Phi} & =\frac{1}{A}\Phi'-\frac{B}{A}\Phi,\\
\tilde{\psi} & =\frac{1}{C}\psi'-\frac{D}{C}\psi.
\end{cases}
\end{equation}
Moreover, it will be useful to define
\begin{equation}
\tilde{A}=2\bar{u}A,\quad\tilde{B}=\bar{u}'+\left(2B-v\right)\bar{u},\quad\tilde{C}={\displaystyle -C,}\quad\tilde{D}=3\tilde{v}+v-D,
\end{equation}
so that
\begin{equation}
\Theta=\tilde{A}\tilde{\Phi}+\tilde{B}\Phi+\tilde{C}\tilde{\psi}+\tilde{D}\psi.\label{eq:3.Ttildes}
\end{equation}
Now, if we insert this version of $\Theta$ into (\ref{eq:3.S2preABCD})
and rewrite, as much as possible, the remaining terms that contain
$\Phi'$ and $\psi'$ in terms of $\tilde{\Phi}$ and $\tilde{\psi}$,
we can simplify $S^{\left(2\right)}$ into a form that contains only
the $\tilde{\Phi}$ and $\tilde{\psi}$ perturbation variables plus
four terms that do not: a $\Phi^{2}$ term, a $\psi^{2}$ term, a
$\Phi\psi$ term and a $\Phi'\psi$ term. The coefficients of these
four terms, constructed out of the four functions $A$, $B$, $C$,
and $D$ and their derivatives, we can then \textsl{choose} to set
to zero. In other words, provided $A$, $B$, $C$, and $D$ are \textsl{any
}solutions satisfying the set of four ODEs resulting from the demand
that these four coefficients vanish, then all terms involving $\Phi$
and $\psi$ can be eliminated --- meaning that we can write $S^{\left(2\right)}$
purely in terms of the new fields $\tilde{\Phi}$ and $\tilde{\psi}$.
The result is:
\begin{equation}
S^{\left(2\right)}=\varepsilon^{2}\int\mathrm{d}^{4}x\,\frac{a^{2}}{2}\Big\{\tilde{u}\tilde{A}^{2}\tilde{\Phi}'^{2}+\tilde{u}\tilde{C}^{2}\tilde{\psi}'^{2}+2\tilde{u}\tilde{A}\tilde{C}\tilde{\Phi}'\tilde{\psi}'+p\tilde{\Phi}'\tilde{\psi}+2q\tilde{\Phi}\tilde{\psi}+m_{1}\tilde{\Phi}^{2}+m_{2}\tilde{\psi}^{2}\Big\},\label{eq:S2tildes}
\end{equation}
where the coefficients $p$, $q$, $m_{1}$ and $m_{2}$ are given
in terms of $A$, $B$, $C$, and $D$ in Appendix \ref{a:Coefficients}.
Moreover, the full set of ODEs that the latter need to satisfy are
also given there. They are highly nontrivial, and we make no attempt
to solve them --- but they do, at the very least, imply that solutions
for $A$, $B$, $C$, and $D$ such that the second order action acquires
the desired canonical form (\ref{eq:S2tildes}) exist.

The point can here be raised that it may appear as though we've eliminated
degrees of freedom in going from the second order action in terms
of $\Phi$ and $\psi$ to its canonical version in terms of $\tilde{\Phi}$
and $\tilde{\psi}$ --- specifically, those contained in the squared
double derivative terms. We can, rather, think of these degrees of
freedom as having been absorbed into the time dependent functions
$A$, $B$, $C$, and $D$. One way to look at this is as follows:
If we had kept $S^{\left(2\right)}$ in terms of $\Phi$ and $\psi$,
we would end up with equations of motion in the form of third-order
ODEs, whose solutions could in principle be obtained after the background
is solved for. Instead, we redefined everything in terms of the new
fields $\tilde{\Phi}$ and $\tilde{\psi}$, which (due to the now
canonical form of $S^{\left(2\right)}$) obey \textsl{second-order
}ODEs. Yet, this can be achieved only on the condition that the functions
$A$, $B$, $C$, and $D$ permitting this redefinition themselves
obey a set of ODEs (that given in Appendix \ref{a:Coefficients})
which, indeed, turn out to be third-order differential equations ---
the solutions for which, again, depend on the background.

Next, we diagonalize the kinetic term. We present the details of the
procedure in Appendix \ref{a:Diagonalization}, and here simply state
the result. Performing a field redefinition
\begin{equation}
\begin{cases}
\zeta & =\tilde{A}\tilde{\Phi}+\tilde{C}\tilde{\psi},\\
\xi & =-\tilde{C}\tilde{\Phi}+\tilde{A}\tilde{\psi},
\end{cases}
\end{equation}
the second order action (\ref{eq:S2tildes}) becomes:
\begin{equation}
S^{\left(2\right)}=\varepsilon^{2}\int\mathrm{d}^{4}x\,\frac{a^{2}}{2}\Big\{\tilde{u}\zeta'^{2}+c_{1}\zeta^{2}+c_{2}\xi^{2}+c_{3}\zeta\xi+c_{4}\zeta'\xi\Big\},\label{3.eq:S2zx}
\end{equation}
with $c_{1}$, $c_{2}$, $c_{3}$ and $c_{4}$ given explicitly in
Appendix \ref{a:Diagonalization}. Thus, we see that there is in fact
\textsl{only one} propagating degree of freedom, namely the field
$\zeta$.

We can simplify this further still, by solving for the non-propagating
field $\xi$ and substituting it back into the action. Its equation
of motion is found by varying (\ref{3.eq:S2zx}) with respect to $\xi$
and simply yields:
\begin{equation}
\xi=-\frac{1}{2c_{2}}\left(c_{3}\zeta+c_{4}\zeta'\right).
\end{equation}
Inserting this back into (\ref{3.eq:S2zx}), expanding, and integrating
by parts, we obtain a remarkably simple (modulo coefficient determination)
second order action:
\begin{equation}
S^{\left(2\right)}=\varepsilon^{2}\int\mathrm{d}^{4}x\,\frac{a^{2}}{2}\Big\{ C_{1}\zeta'^{2}+C_{2}\zeta^{2}\Big\},\label{eq:3.S2z}
\end{equation}
where
\begin{align}
C_{1}=\: & \tilde{u}-\frac{c_{4}^{2}}{4c_{2}},\\
C_{2}=\: & c_{1}+\frac{c_{3}}{4c_{2}}\left[2\mathcal{H}c_{4}-c_{3}\right]+\left(\frac{c_{3}c_{4}}{4c_{2}}\right)'.
\end{align}

Unfortunately, as an exact analytical computation of $C_{1}$ and
$C_{2}$ seems to be (at least with our current approach) an immensely
nontrivial problem, (\ref{eq:3.S2z}) as such can only give us the
form of the equation of motion for perturbations in this theory, but
not an indication as to whether or not solutions thereof are in general
stable. In other words, we cannot immediately conclude from (\ref{eq:3.S2z})
if $\zeta$ is ghost-like nor tachyonic. Instead, we shall have to
resort to approximations, which we turn to in the next section.

\section{Stability Analysis\label{sec:Stability-Analysis}}

The first step in determining whether cosmological solutions of this
theory are perturbatively stable is to ask where we would expect things
to go wrong. We can infer right away that we need not be concerned
when they pass through the attractive gravity region $\mathbb{G}_{+}$
of field space. The reason is simply that a gauge choice (in the form
of a judicious reparametrization of the matter fields, as described
in Appendix \ref{a:Consistency}) can here be made to reduce the theory
to ordinary (attractive) Einstein gravity. Therefore, the regime where
we might worry about the possible appearance of perturbative ghosts
(or tachyons) is the repulsive gravity region $\mathbb{G}_{-}$ ---
this being, indeed, where all the novel effects of the theory come
in.

In particular, we will focus our attention on the evolution of perturbations
around the area of $\mathbb{G}_{-}$ where the background dilaton
becomes very small, while the background Higgs does not. In other
words, we will consider the limit in which $\phi_{0}\rightarrow0$,
but $s_{0}$ as well as their first derivatives remain finite. This
corresponds to the regime in field space where the Newton constant
attains its \textsl{most negative} value, and so for this reason we
refer to it as the \textsl{deep antigravity }limit.

We will base our overall determination of the stability of perturbations
in this theory upon their behaviour in this limit, for two reasons.
First, it is a limit which \textsl{all} background solutions passing
through $\mathbb{G}_{-}$ necessarily exhibit at some point (irrespective
of whether they contain anisotropies, which we have not even included
in our analysis here) and which, as we shall see in the next subsection,
can be taken via a more or less straightforward procedure. Second,
though a verdict on stability/instability in the deep antigravity
regime need not translate to the same throughout \textsl{all} of $\mathbb{G}_{-}$,
we can offer an argument for why it is nonetheless sufficiently indicative
for allowing us to draw a conclusion on the entire theory. In the
case that a ghost/tachyon \textsl{is} found here, we would immediately
deduce that the theory is problematic (or, at best, contingent upon
the instability having a very short duration). In the case that it
is \textsl{not} found, we are justified in asserting that the theory
is safe. The reason for this is as follows: Even if a perturbative
instability actually does develop at any \textsl{other} point during
the passage through $\mathbb{G}_{-}$ (e.g. at the boundary between
$\mathbb{G}_{+}$ and $\mathbb{G}_{-}$), it would have to be so short-lived
as to be quickly back under control by the (ghost-free) deep antigravity
regime roughly corresponding to the half-way point of the (already,
generically brief) repulsive gravity phase. In other words, even if
the kinetic term (or mass term) reversed signs anywhere else during
$\mathbb{G}_{-}$ but went back to the correct sign about half-way
through it, this potential momentary instability may not be regarded
as so catastrophic as to jeopardize the entire theory.

In summary: To establish the stability of perturbations in this theory,
it will suffice to take the deep antigravity limit of the second-order
action, i.e. $\phi_{0}\rightarrow0$ while keeping all other background
quantities finite.

The second step in addressing the issue of stability is to ask which
of the perturbation variables we should watch out for. Seeing as one
of the two matter fields (namely the dilaton) is what appears as a
(potentially problematic) ghost in the full action, it is the matter
perturbation that is the sole concern here (while the metric perturbation's
stability should not have to worry us). In other words, in lieu of
determining whether the single perturbation variable $\zeta$ is stable,
it will suffice to try to find whether $\psi$ is stable. Of course,
taking the deep antigravity limit in the second-order action (\ref{eq:3.S2z})
for the former is made difficult by the fact that we have no analytic
expression --- nor, indeed, even a good ansatz --- for the functions
$A$, $B$, $C$ and $D$ upon which the coefficients appearing therein
ultimately depend. But we \textsl{can} take this limit in the case
of the latter --- in particular, by looking at the last version of
the second-order action we obtained just before these four functions
made their appearance.

\subsection{Deep Antigravity Approximation}

Consider $S^{\left(2\right)}$ in the form (\ref{eq:3.S2preABCD})
just prior to making the field redefinitions to remove the squared
double derivative terms. We rewrite it here for convenience:
\begin{align}
S^{\left(2\right)}=\varepsilon^{2}\int\mathrm{d}^{4}x\,\frac{a^{2}}{2}\Big\{ & \tilde{u}\Theta'^{2}+\hat{u}\Theta^{2}-3\bar{v}\tilde{u}'\psi\Theta+\hat{v}\psi'\Theta+3\tilde{u}'\psi'\Theta'-9u'^{2}\psi'^{2}+9\bar{v}u'^{2}\psi^{2}\nonumber \\
 & +\frac{5}{3}z\Phi'^{2}-2\phi_{0}u'\Phi'\left(\psi'-v\psi\right)-24\phi_{0}u'\left[\tilde{v}'+\tilde{v}^{2}\right]\Phi\psi\Big\},\label{eq:S2preABCD2}
\end{align}
where $\Theta$ should be everywhere understood according to (\ref{eq:P&T})
as written in terms of the metric perturbation $\Phi$ and matter
perturbation $\psi$, namely:
\begin{equation}
\Theta=2\bar{u}\Phi'+\left(\bar{u}'-v\bar{u}\right)\Phi-\psi'+\left(3\tilde{v}+v\right)\psi.\label{eq:T2}
\end{equation}
The advantage of taking the $\phi_{0}\rightarrow0$ limit of (\ref{eq:S2preABCD2})
(as opposed to the second-order action (\ref{eq:3.S2z}) for $\zeta$)
is that all of the coefficients are known exactly. The strategy then
is to insert their full expressions and collect the result in powers
of $\phi_{0}$. The lowest order (or highest inverse order) terms
will be the dominant ones, and therefore the ones of interest.

Now, extracting their explicit forms from Table \ref{tab:2}, we note
that all of the coefficient functions appearing in (\ref{eq:S2preABCD2})
with (\ref{eq:T2}) substituted are in fact exact (finite) series
in powers of $\phi_{0}$, with the sole exception of the function
$\bar{u}$. Taylor expanding the latter in $\phi_{0}$, inserting
the exact expressions for the rest of the coefficients, collecting
the integrand in powers of $\phi_{0}$, integrating by parts and simplifying
as much as possible, we get:
\begin{equation}
S^{\left(2\right)}=\varepsilon^{2}\int\mathrm{d}^{4}x\,\frac{a^{2}}{2}\left\{ \left[-\frac{24\phi_{0}'^{4}s_{0}^{2}}{\phi_{0}^{6}}+\cdots\right]\psi^{2}+\frac{3\phi_{0}'^{2}s_{0}^{2}}{\phi_{0}^{4}}\psi'^{2}+\cdots\right\} ,\label{eq:4.S2approx}
\end{equation}
where $\cdots$ here refers to higher order terms in $\phi_{0}$.
These include all contributions from the $\Phi$ perturbation as well
as the squared double derivative terms, which accordingly do not concern
us.

We thus see that the significant contributions to the second order
action in the deep antigravity limit are simply a kinetic term and
a mass term for the matter perturbation variable $\psi$. If we write
(\ref{eq:4.S2approx}) more suggestively (dropping the higher order
terms) as
\begin{equation}
S^{\left(2\right)}=\varepsilon^{2}\int\mathrm{d}^{4}x\,\left(\frac{\sqrt{3}a\phi_{0}'s_{0}}{\sqrt{2}\phi_{0}^{2}}\right)^{2}\left\{ \psi'^{2}-\left(2\sqrt{2}\frac{\phi_{0}'}{\phi_{0}}\right)^{2}\psi^{2}\right\} ,
\end{equation}
we see that $\psi$ has the correct (positive) sign for its kinetic
term, and also the correct (negative) sign for its mass term. This
means, respectively, that it is neither a ghost nor a tachyon. Hence,
based on our entire discussion in this section up to now, we conclude
that cosmological solutions of this theory are perturbatively stable,
and therefore viable.

\subsection{Equation of Motion and Solution}

Finally, we derive the equation of motion for the perturbation in
this limit, and obtain its solution. Varying the approximated second-order
action (\ref{eq:4.S2approx}) with respect to $\psi$ yields:
\begin{equation}
0=\psi''+\left(-4\frac{\phi_{0}'}{\phi_{0}}+\cdots\right)\psi'+\left(2\sqrt{2}\frac{\phi_{0}'}{\phi_{0}}+\cdots\right)^{2}\psi.\label{eq:4.PEOM}
\end{equation}
To solve this, we can insert the background solution for $\phi_{0}$
obtained in \cite{Bars2012a}:
\begin{equation}
\phi_{0}\left(\eta\right)=\mathscr{A}\frac{{\rm sn}\left(\eta,m\right)}{{\rm dn}\left(\eta,m\right)},\label{eq:4.p0sol}
\end{equation}
where $\eta=(\tau+\tau_{0})/T$ is a shifted and rescaled time coordinate,
$\mathscr{A}$ is a constant, and ${\rm sn}$ and ${\rm dn}$ are
Jacobi elliptic functions, with the parameter $m=1/2$.

To get a better intuition for (\ref{eq:4.p0sol}), we can look at
its Taylor series in $\eta$: 
\begin{equation}
\phi_{0}=\mathscr{A}\left[\eta-\frac{1}{12}\eta^{3}-\frac{1}{60}\eta^{5}+\mathcal{O}\left(\eta^{7}\right)\right].
\end{equation}
So we see that the $\phi_{0}\rightarrow0$ limit corresponds to $\eta\rightarrow0$.
Thus, we can use the approximation $\phi_{0}\approx\mathscr{A}\eta$
in this regime.

At this point we can take two possible approaches towards solving
(\ref{eq:4.PEOM}). One is to directly insert (\ref{eq:4.p0sol})
into it, as is. The resulting ODE actually turns out to have an exact
solution. However, its form is not particularly illuminating; it is
given in terms of Jacobi elliptic functions and the Heun general function.
For the sake of completeness, we write it down in full in Appendix
\ref{a:Solution}.

\begin{figure}[h]
\includegraphics[scale=0.5]{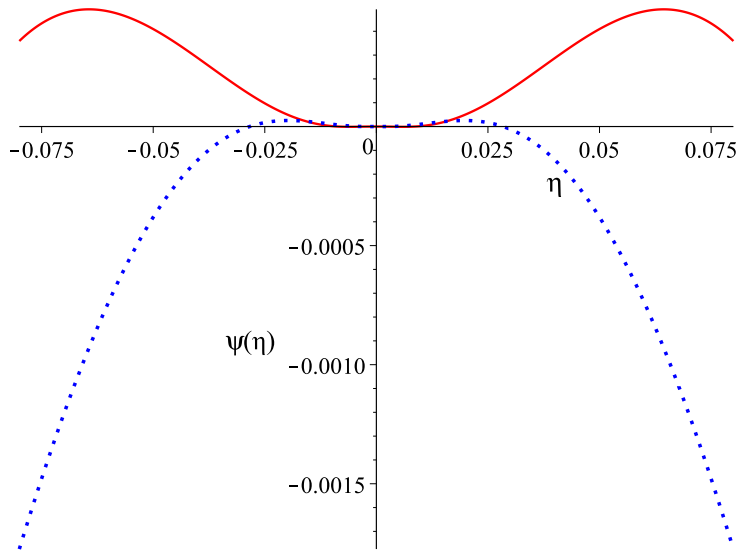}\caption{\label{fig:PerturbativeSol}The two independent solutions for the
matter perturbation in the deep antigravity regime: $\psi_{1}$ (solid
red) and $\psi_{2}$ (dotted blue).}

\end{figure}

A second, and perhaps more instructive approach to solving (\ref{eq:4.PEOM})
is to make some further approximations first. As suggested, we can
try to use $\phi_{0}\approx\mathscr{A}\eta$, which entails $({\rm d}\phi_{0}/{\rm d}\eta)/\phi_{0}\approx1/\eta.$
With this, our ODE for $\psi$, written in terms of the variable $\eta$,
becomes:
\begin{equation}
0=\frac{{\rm d}^{2}\psi}{{\rm d}\eta^{2}}-\frac{4}{\eta}\frac{{\rm d}\psi}{{\rm d}\eta}+\frac{8}{\eta^{2}}\psi,
\end{equation}
which is easily solvable. Its two independent solutions are:
\begin{equation}
\begin{cases}
\psi_{1} & =\left|\eta\right|^{5/2}\sin\left(\frac{\sqrt{7}}{2}\ln\left|\eta\right|\right),\\
\psi_{2} & =\left|\eta\right|^{5/2}\cos\left(\frac{\sqrt{7}}{2}\ln\left|\eta\right|\right).
\end{cases}
\end{equation}
We plot each of these in Figure \ref{fig:PerturbativeSol}.

The general solution is then a superposition of the two:
\begin{equation}
\psi=\left|\eta\right|^{5/2}\left[\mathscr{C}_{1}\sin\left(\frac{\sqrt{7}}{2}\ln\left|\eta\right|\right)+\mathscr{C}_{2}\cos\left(\frac{\sqrt{7}}{2}\ln\left|\eta\right|\right)\right],
\end{equation}
where $\mathscr{C}_{1}$ and $\mathscr{C}_{2}$ are constants.

We remark that, for all solutions, $\psi\rightarrow0$ as $\eta\rightarrow0$.
That is to say, perturbations become suppressed in the deep antigravity
regime. This behaviour is markedly contrary to what happens in ordinary
(Einstein) gravity. There, it is precisely the fact that gravity is
\textsl{attractive} which leads to the growth of inhomogeneities,
and consequently, structure formation: In brief, overdensities increase
because their surrounding matter is further attracted into them \cite{Mukhanov1992,Brandenberger2004}.
When gravity is strongly \textsl{repulsive}, on the other hand, what
may start out as an overdensity will soon be repelled apart. What
is more, we see here that when gravity is as repulsive as it can possibly
be, matter becomes evenly repelled to the point that any prior fluctuation
will completely vanish, leaving us with an entirely homogeneous Universe.
This suggests that this theory may be able to supply its own resolution
to the problem of why, indeed, we observe our own Universe to be so
nearly homogeneous today.

\section{Conclusions\label{sec:Conclusions}}

\subsection{Summary of Results}

In this paper, we have considered the Weyl-symmetric theory (\ref{eq:S})
of two scalar fields, the dilaton $\phi$ and the Higgs $s$, conformally
coupled with opposite signs to Einstein gravity, and we have shown
that its isotropic cosmological solutions admit scalar perturbations
whose second-order action (\ref{eq:3.S2z}) can in general be written
in a canonical form,
\begin{equation}
S^{\left(2\right)}=\varepsilon^{2}\int\mathrm{d}^{4}x\,\frac{a^{2}}{2}\Big\{ C_{1}\zeta'^{2}+C_{2}\zeta^{2}\Big\},
\end{equation}
in terms of a single propagating perturbation variable $\zeta$, which
is a linear combination of the metric and matter perturbations, and
their first derivatives.

Moreover, we have studied this theory in the regime where the dilaton
approaches zero but the Higgs remains finite (corresponding to the
epoch of strongly repulsive gravity roughly mid-way between a Big
Crunch and a Big Bang in its cosmological solutions), and found that
the second-order action (\ref{eq:4.S2approx}) there, which can be
approximated purely in terms of the single matter perturbation $\psi$,
\begin{equation}
S^{\left(2\right)}=\varepsilon^{2}\int\mathrm{d}^{4}x\,\frac{a^{2}}{2}\left(\frac{\sqrt{3}\phi_{0}'s_{0}}{\phi_{0}^{2}}\right)^{2}\left\{ \psi'^{2}-\left(2\sqrt{2}\frac{\phi_{0}'}{\phi_{0}}\right)^{2}\psi^{2}\right\} ,
\end{equation}
contains neither ghosts nor tachyons. These results indicate that
perturbations in this theory are stable, and therefore its cosmological
solutions are viable. Furthermore, we have explicitly solved for the
evolution of the matter perturbation in this regime and found it to
be completely suppressed when the dilaton vanishes (i.e. when gravity
is maximally repulsive), hinting at a solution to the homogeneity
problem.

\subsection{Future Work}

Here we have presented the most basic possible analysis of perturbations
in cosmological solutions of this theory. However, there is a wealth
of more detailed aspects of these solutions which we have neglected
for the sake of simplicity, and which offer many potentially fruitful
avenues of future investigation.

First, there is the fact that we have only been working with the action
(\ref{eq:S}) with a vanishing potential $V\left(\phi,s\right)$, and without
any radiative matter. Both the potential and radiative matter are important
for the background cosmology which the authors of
\cite{Bars2010,Bars2011a,Bars2011b,Bars2011c,Bars2012a,Bars2012b,Bars2013a,Bars2013b,Bars:2013c}
have in mind and hence
further insight may be gained by repeating the procedure in this
paper with $V$ kept as an arbitrary function and/or with simple choices
of interactions; a conformally-extended symmetry-breaking-type potential
of the form 
\[
V=\frac{\lambda}{4}\left(\frac{s^{2}}{2}-\alpha^{2}\phi^{2}\right)^{2}+\frac{\lambda'}{4}\phi^{4},
\]
(motivated from \cite{Bars2013a}) would be one of the first obvious
options to try --- with $\lambda$, $\lambda'$, and $\alpha$ here
denoting coupling constants.

Second, we have not considered anisotropies in our analysis. Yet,
as mentioned in the introduction, the inclusion of anisotropies is
what leads to a (potential-independent) attractor mechanism which
forces cosmological solutions of this theory through the origin of
field space (with a loop in the $\mathbb{G}_{-}$ region between Crunch
and Bang, illustrated by the red curve in Figure \ref{fig:BackgroundSol}).
Moreover, the total suppression of perturbations (portrayed in Figure
\ref{fig:PerturbativeSol}) about mid-way through the (repulsive gravity)
period between the (attractive gravity) contraction and expansion
epochs of the Universe in this theory appears to be novel; in other
``matter bounce'' scenarios this is not the case --- but, usually
because of anisotropies. These two reasons together offer strong motivation
for expanding the present analysis to include anisotropies, and thereby
determine precisely to what extent they play a role at the perturbative
level.

Third, we have only considered the evolution of scalar metric perturbations.
In cosmologies based on Einstein gravity, it is the scalar metric sector
in which the strongest instabilities tend to occur. However, in the
background theory we are considering the gravitational wave sector may also
show instabilities which need to be analyzed. Thus, it will be important to study
the behaviour of the other two classes of metric perturbations, namely
vector and tensor perturbations, before drawing
definite conclusions about the stability of the model.
The study of tensor perturbations is of particular
interest since it will yield the predictions of the model for the
spectrum of gravitational waves.

Fourth, our treatment has been a purely classical one. This leaves
the door open for a quantum analysis of this problem.

\section*{Acknowledgements}

This work was supported by the Natural Sciences and Engineering Research
Council of Canada. We would like to thank E. Ferreira and H. Bazrafshan
Moghaddam for helpful discussions. We are particularly grateful to Paul
Steinhardt for his careful reading of the manuscript and for his insightful
comments.

\appendix

\section{Consistency Check\label{a:Consistency}}

We can eliminate one of the two scalar field degrees of freedom via
the following reparametrizations of the dilaton and the Higgs, respectively:
\begin{equation}
\begin{cases}
\phi_{\mathrm{E}} & =\sqrt{6}\cosh\left(\sigma/\sqrt{6}\right),\\
s_{\mathrm{E}} & =\sqrt{6}\sinh\left(\sigma/\sqrt{6}\right),
\end{cases}\label{eq:phi&sE}
\end{equation}
in terms of a single scalar field $\sigma$. Setting $\phi=\phi_{{\rm E}}$
and $s=s_{{\rm E}}$, the Weyl-symmetric theory (\ref{eq:S}) reduces
to ordinary Einstein gravity minimally coupled to this single scalar
field,
\begin{equation}
S_{\mathrm{E}}\left[\sigma;g_{\mu\nu}\right]=\int\mathrm{d}^{4}x\sqrt{-g}\left\{ -\frac{1}{2}R+\frac{1}{2}g^{\mu\nu}\nabla_{\mu}\sigma\nabla_{\nu}\sigma\right\} .\label{eq:SE}
\end{equation}

Moreover, setting $\phi=\phi_{{\rm E}}$ and $s=s_{{\rm E}}$ in the
gravitational (\ref{eq:Gravity}) and matter (\ref{eq:Matter}) equations
of motion reduces them, as expected, to the usual Einstein equation
and the massless Klein-Gordon equation for the scalar field $\sigma$,
respectively:
\begin{eqnarray}
 & G_{\alpha\beta}=-T_{\alpha\beta},\label{eq:Einstein}\\
 & g^{\mu\nu}\nabla_{\mu}\nabla_{\nu}\sigma=0.\label{eq:KG}
\end{eqnarray}

Now, in order to recover the results of standard cosmological perturbation
theory in Einstein gravity --- where the perturbed metric (\ref{eq:g})
is the same --- we need to choose appropriate reparametrizations of
$\phi_{0}$, $\Pi$, $s_{0}$, and $\Theta$ such that, in this limit,
the full dilaton and Higgs solutions become $\phi=\phi_{\mathrm{E}}$
and $s=s_{\mathrm{E}}$. Moreover, the scalar field $\sigma$ in terms
of which $\phi_{{\rm E}}$ and $s_{{\rm E}}$ are written (\ref{eq:phi&sE})
should itself decompose, as usual, into a homogeneous (purely time-dependent)
part $\sigma_{0}\left(\tau\right)$ plus a (spacetime-dependent) perturbation
$\Delta\left(\tau,\mathbf{x}\right)$, i.e.
\begin{equation}
\sigma=\sigma_{0}+\varepsilon\Delta.\label{eq:sigma}
\end{equation}
The way to achieve this is to Taylor expand $\phi_{\mathrm{E}}=\sqrt{6}\cosh(\sigma/\sqrt{6})$
and $s_{\mathrm{E}}=\sqrt{6}\sinh(\sigma/\sqrt{6})$ in $\sigma$
about the background value $\sigma_{0}$, and match the resulting
series, in powers of $\varepsilon$, with $\phi=\phi_{0}+\varepsilon\Pi$
and $s=s_{0}+\varepsilon\Theta$, respectively. This entails that
the appropriate reparametrizations that recover the Einstein limit
for the background fields are simply
\begin{equation}
\begin{cases}
\phi_{0\mathrm{E}} & =\sqrt{6}\cosh\left(\sigma_{0}/\sqrt{6}\right),\\
s_{0\mathrm{E}} & =\sqrt{6}\sinh\left(\sigma_{0}/\sqrt{6}\right),
\end{cases}\label{eq:phi&s0E}
\end{equation}
while for their respective perturbations, they must be
\begin{align}
\Pi_{\mathrm{E}} & =\sum_{n=0}^{\infty}\frac{\sinh(\sigma_{0}/\sqrt{6})}{6^{n}\left(2n+1\right)!}\left[1+\frac{\coth(\sigma_{0}/\sqrt{6})}{\sqrt{6}\left(2n+2\right)}\varepsilon\Delta\right]\varepsilon^{2n}\Delta^{2n+1}\label{eq:3.PIEseries}\\
 & =\sinh\left(\frac{\sigma_{0}}{\sqrt{6}}\right)\Delta+\frac{1}{2\sqrt{6}}\cosh\left(\frac{\sigma_{0}}{\sqrt{6}}\right)\varepsilon\Delta^{2}+\mathcal{O}\left(\varepsilon^{2}\right),\label{eq:3.PIE}
\end{align}
and
\begin{align}
\Theta_{\mathrm{E}} & =\sum_{n=0}^{\infty}\frac{\cosh(\sigma_{0}/\sqrt{6})}{6^{n}\left(2n+1\right)!}\left[1+\frac{\tanh(\sigma_{0}/\sqrt{6})}{\sqrt{6}\left(2n+2\right)}\varepsilon\Delta\right]\varepsilon^{2n}\Delta^{2n+1}\label{eq:3.ThetaEseries}\\
 & =\cosh\left(\frac{\sigma_{0}}{\sqrt{6}}\right)\Delta+\frac{1}{2\sqrt{6}}\sinh\left(\frac{\sigma_{0}}{\sqrt{6}}\right)\varepsilon\Delta^{2}+\mathcal{O}\left(\varepsilon^{2}\right).\label{eq:3.ThetaE}
\end{align}

Upon substituting $\phi_{0}=\phi_{0{\rm E}}$, $\Pi=\Pi_{{\rm E}}$,
$s_{0}=s_{0{\rm E}}$ and $\Theta=\Theta_{{\rm E}}$, which turn (\ref{eq:z})
into
\begin{equation}
\begin{cases}
z_{\mathrm{E}} & =6,\\
\tilde{z}_{\mathrm{E}} & =-\sigma_{0}'^{2},
\end{cases}
\end{equation}
and (\ref{eq:a&U}) into
\begin{equation}
\begin{cases}
\alpha_{\mathrm{E}} & =\varepsilon\frac{1}{2}\Delta^{2}+...,\\
\tilde{\alpha}_{\mathrm{E}} & =-\sigma_{0}'\Delta'-\varepsilon\frac{1}{12}\sigma_{0}'^{2}\Delta^{2}+...,
\end{cases}\quad{\rm and}\quad\begin{cases}
\Upsilon_{{\rm E}} & =6\Phi-\varepsilon\frac{1}{2}\Delta^{2}+...,\\
\tilde{\Upsilon}_{{\rm E}} & =-\sigma_{0}'^{2}\Phi+\sigma_{0}'\Delta'+\varepsilon\frac{1}{12}\sigma_{0}'^{2}\Delta^{2}+...,
\end{cases}\label{eq:a&UE}
\end{equation}
where $\cdots$ denotes higher order terms in $\varepsilon$, all
of the equations of motion given in Table \ref{tab:1} reduce to their
appropriate counterparts in the usual Einstein case \cite{Mukhanov1992},
given in Table \ref{tab:a}.

\begin{table}[h]
\begin{tabular}{|>{\centering}m{2.5cm}||>{\centering}m{3.25cm}|>{\centering}m{7.5cm}|}
\hline 
\multirow{2}{2.5cm}{EOM} & \multirow{2}{3.25cm}{~$\mathcal{O}\left(1\right)$} & \multirow{2}{7.5cm}{~$\mathcal{O}\left(\varepsilon\right)$}\tabularnewline
 &  & \tabularnewline
\hline 
\hline 
\multirow{2}{2.5cm}{(\ref{eq:Einstein}) time-time} & \multirow{2}{3.25cm}{~$0=\mathcal{H}^{2}-\frac{1}{6}\sigma_{0}'^{2}$} & \multirow{2}{7.5cm}{~$0=\nabla^{2}\Phi-3\mathcal{H}\left(\Phi'+\mathcal{H}\Phi\right)+\frac{1}{2}\sigma_{0}'^{2}\Phi-\frac{1}{2}\sigma_{0}'\Delta'$}\tabularnewline
 &  & \tabularnewline
\hline 
\multirow{2}{2.5cm}{(\ref{eq:Einstein}) space-space} & \multirow{2}{3.25cm}{~$0=2\mathcal{H}'+\mathcal{H}^{2}+\frac{1}{2}\sigma_{0}'^{2}$} & \multirow{2}{7.5cm}{~$0=\Phi''+3\mathcal{H}\Phi'+\left(2\mathcal{H}'+\mathcal{H}^{2}\right)\Phi+\frac{1}{2}\sigma_{0}'^{2}\Phi-\frac{1}{2}\sigma_{0}'\Delta'$}\tabularnewline
 &  & \tabularnewline
\hline 
\multirow{2}{2.5cm}{(\ref{eq:Einstein}) time-space} & \multirow{2}{3.25cm}{~Trivial.} & \multirow{2}{7.5cm}{~$0=\partial_{i}\left(\mathcal{H}\Phi+\Phi'-\frac{1}{2}\sigma_{0}'\Delta\right)$}\tabularnewline
 &  & \tabularnewline
\hline 
\multirow{2}{2.5cm}{(\ref{eq:KG})} & \multirow{2}{3.25cm}{~$0=\sigma_{0}''+2\mathcal{H}\sigma_{0}'$} & \multirow{2}{7.5cm}{~$0=\Delta''+2\mathcal{H}\Delta'-\nabla^{2}\Delta-4\sigma_{0}'\Phi'$}\tabularnewline
 &  & \tabularnewline
\hline 
\end{tabular}\caption{\label{tab:a}Perturbed equations of motion in Einstein gravity.}

\end{table}

Moreover, the second order action (\ref{eq:S2}) reduces correctly
\cite{Mukhanov1992} in the Einstein limit:
\begin{equation}
S_{{\rm E}}^{\left(2\right)}=\varepsilon^{2}\int\mathrm{d}^{4}x\,\frac{a^{2}}{2}\bigg\{-6\left[\Phi'^{2}+\frac{1}{3}\left(\nabla\Phi\right)^{2}\right]+\Delta'^{2}-\left(\nabla\Delta\right)^{2}+8\sigma_{0}'\Phi'\Delta\bigg\}.\label{eq:S2E}
\end{equation}
Because of the $\mathcal{O}\left(\varepsilon\right)$ terms appearing
in $\Pi_{{\rm E}}$ and $\Theta_{{\rm E}}$, one may worry about possible
$\mathcal{O}(\varepsilon^{2})$ contributions to this limit from the
first order action, which can be simplified to:
\begin{equation}
S^{\left(1\right)}=\varepsilon\int\mathrm{d}^{4}x\,\frac{a^{2}}{2}\bigg\{-z\Phi''+2\left[\left(\mathcal{H}^{2}+\mathcal{H}'\right)\alpha-\tilde{\alpha}\right]\bigg\}.\label{eq:S1}
\end{equation}
Indeed, the $\alpha$ and $\tilde{\alpha}$ terms do contain $\mathcal{O}\left(\varepsilon\right)$
contributions in the Einstein limit as per (\ref{eq:a&UE}), leading
a priori to $\mathcal{O}(\varepsilon^{2})$ additions to (\ref{eq:S2E})
from (\ref{eq:S1}). However, it is possible to verify that the entire
term appearing in the square brackets in (\ref{eq:S1}) actually vanishes
at $\mathcal{O}\left(\varepsilon\right)$ in the Einstein limit by
virtue of the background equations.

\section{Perturbed Action Coefficients\label{a:Coefficients}}

For convenience, we define:
\begin{equation}
\overset{*}{u}=\bar{u}'-v\bar{u},
\end{equation}
as well as:
\begin{align}
\bar{A}=\: & \tilde{A}'+A\tilde{B},\\
\bar{B}=\: & \tilde{B}'+B\tilde{B},\\
\bar{C}=\: & \tilde{C}'+C\tilde{D},\\
\bar{D}=\: & \tilde{D}'+D\tilde{D},\\
\overset{*}{D}=\: & -6\mathcal{H}\tilde{u}'D-3\left(\tilde{u}'D\right)',\\
\mathfrak{D}_{1}=\: & 2\bar{u}\hat{v}-6\bar{u}\tilde{u}'D-2\phi_{0}u',\\
\mathfrak{D}_{2}=\: & 3\tilde{u}'D-\hat{v}-9u'^{2},\\
E=\: & \tilde{u}\left[\bar{A}^{2}-2\tilde{A}\left(\mathcal{H}\bar{A}+A\bar{B}\right)\right]-\left(\tilde{u}\tilde{A}\bar{A}\right)',\\
F=\: & \tilde{u}\left[\bar{C}^{2}-2\tilde{C}\left(\mathcal{H}\bar{C}+C\bar{D}\right)\right]-\left(\tilde{u}\tilde{C}\bar{C}\right)',\\
G=\: & 2\tilde{u}\left[\left(\bar{A}\bar{C}-C\tilde{A}\bar{D}\right)-\tilde{C}\left(2\mathcal{H}\bar{A}+A\bar{B}\right)\right]-2\left(\tilde{u}\bar{A}\tilde{C}\right)',\\
I=\: & 2\tilde{u}\bar{B}\left[\bar{A}-\left(2\mathcal{H}+B\right)\tilde{A}\right]-2\left(\tilde{u}\tilde{A}\bar{B}\right)',\\
J=\: & 2\tilde{u}\bar{D}\left[\bar{A}-\left(2\mathcal{H}+D\right)\tilde{A}\right]-2\left(\tilde{u}\tilde{A}\bar{D}\right)',\\
K=\: & 2\tilde{u}\bar{B}\left[\bar{C}-\left(2\mathcal{H}+B\right)\tilde{C}\right]-2\left(\tilde{u}\tilde{C}\bar{B}\right)',\\
L=\: & 2\tilde{u}\bar{D}\left[\bar{C}-\left(2\mathcal{H}+D\right)\tilde{C}\right]-2\left(\tilde{u}\tilde{C}\bar{D}\right)',
\end{align}
Inserting $\Theta$ as given by (\ref{eq:3.Ttildes}) into (\ref{eq:3.S2preABCD})
we get, after simplifications:
\begin{align}
S^{\left(2\right)}=\varepsilon^{2}\int\mathrm{d}^{4}x\,\frac{a^{2}}{2}\Big\{ & \tilde{u}\tilde{A}^{2}\tilde{\Phi}'^{2}+\tilde{u}\tilde{C}^{2}\tilde{\psi}'^{2}+2\tilde{u}\tilde{A}\tilde{C}\tilde{\Phi}'\tilde{\psi}'+p\tilde{\Phi}'\tilde{\psi}+2q\tilde{\Phi}\tilde{\psi}+m_{1}\tilde{\Phi}^{2}+m_{2}\tilde{\psi}^{2}\nonumber \\
 & +\kappa_{1}\Phi'\psi+\kappa_{2}\Phi^{2}+\kappa_{3}\psi^{2}+\kappa_{4}\Phi\psi\Big\},
\end{align}
where
\begin{align}
p=\: & 2\tilde{u}\left(\tilde{A}\bar{C}-\bar{A}\tilde{C}\right)+3\tilde{u}'C\tilde{A},\\
2q=\: & G+2\hat{u}\tilde{A}\tilde{C}+3\tilde{u}'C\bar{A}+AC\mathfrak{D}_{1},\\
m_{1}=\: & E+\hat{u}\tilde{A}^{2}+\frac{5}{3}A^{2}z,\\
m_{2}=\: & F+\hat{u}\tilde{C}^{2}+3\tilde{u}'C\left(\bar{C}+\mathcal{H}C\right)+\frac{3}{2}\left(\tilde{u}'C^{2}\right)'+C^{2}\mathfrak{D}_{2},
\end{align}
and
\begin{align}
\kappa_{1}=\: & \frac{J}{A}-\frac{K}{C}+2\hat{u}\left(\tilde{B}+2\bar{u}\tilde{D}\right)-6\bar{u}\bar{v}\tilde{u}'-\hat{v}\overset{*}{u}+2\bar{u}\overset{*}{D}-3\tilde{u}'\left(\bar{B}-\overset{*}{u}D\right)+2\phi_{0}vu'+\left(D-B\right)\mathfrak{D}_{1},\\
\kappa_{2}=\: & \tilde{u}\bar{B}^{2}-\frac{I}{A}\left(\mathcal{H}+B\right)-\frac{1}{2}\left(\frac{I}{A}\right)'+\hat{u}\tilde{B}\left(\tilde{B}-4\bar{u}\left[\mathcal{H}+B\right]\right)-2\left(\hat{u}\bar{u}\tilde{B}\right)'-\frac{5}{3}\left(\left(B+2\mathcal{H}\right)Bz+\left(Bz\right)'\right),\\
\kappa_{3}=\: & \tilde{u}\bar{D}^{2}-\frac{L}{C}\left(\mathcal{H}+D\right)-\frac{1}{2}\left(\frac{L}{C}\right)'+\hat{u}\tilde{D}\left(\tilde{D}+2\left[\mathcal{H}+D\right]\right)+\left(\hat{u}\tilde{D}\right)'-12\tilde{v}\bar{v}\tilde{u}'-\frac{3}{2}\left(\bar{v}\tilde{u}'\right)'\nonumber \\
 & -\mathcal{H}\hat{v}\left(3\tilde{v}+v\right)-\frac{1}{2}\left(\hat{v}\left[3\tilde{v}+v\right]\right)'+4\tilde{v}\overset{*}{D}-3\tilde{u}'\left(D\bar{D}+\mathcal{H}\left[\tilde{D}'-D^{2}\right]\right)-\frac{1}{2}\left(3\tilde{u}'\left[\tilde{D}'-D^{2}\right]-\overset{*}{D}\right)'\nonumber \\
 & +9\bar{v}u'^{2}-\left(D+2\mathcal{H}\right)D\mathfrak{D}_{2}-\left(D\mathfrak{D}_{2}\right)',\\
\kappa_{4}=\: & 2\tilde{u}\bar{B}\bar{D}-B\frac{J}{A}-\frac{K}{C}\left(2\mathcal{H}+D\right)-\left(\frac{K}{C}\right)'+2\hat{u}\left(\left[\mathcal{H}+4\tilde{v}\right]\tilde{B}-2\bar{u}B\tilde{D}\right)+2\left(\hat{u}\tilde{B}\right)'-3\bar{v}\tilde{u}'\overset{*}{u}\nonumber \\
 & -2\mathcal{H}\hat{v}\overset{*}{u}-\left(\hat{v}\overset{*}{u}\right)'+\overset{*}{u}\overset{*}{D}-3\tilde{u}'\left(D\bar{B}+2\mathcal{H}\left[\bar{B}-\overset{*}{u}D\right]\right)-3\left(\tilde{u}'\left[\bar{B}-\overset{*}{u}D\right]\right)'-24\phi_{0}u'\left(\tilde{v}'+\tilde{v}^{2}\right)\nonumber \\
 & -\left(D+2\mathcal{H}\right)B\mathfrak{D}_{1}-\left(B\mathfrak{D}_{1}\right)'.
\end{align}
Then, the four ODEs that the functions $A$, $B$, $C$, and $D$
need to satisfy in order for $S^{\left(2\right)}$ to be placed in
the form (\ref{eq:S2tildes}) are:
\begin{align}
\kappa_{1}= & 0,\\
\kappa_{2}= & 0,\\
\kappa_{3}= & 0,\\
\kappa_{4}= & 0.
\end{align}

\section{Perturbed Action Diagonalization\label{a:Diagonalization}}

We can write the second order action (\ref{eq:S2tildes}) as:
\begin{align}
S^{\left(2\right)}=\varepsilon^{2}\int\mathrm{d}^{4}x\,\frac{a^{2}}{2}\bigg\{ & \tilde{u}\left(\left[\begin{array}{cc}
\tilde{\Phi}' & \tilde{\psi}'\end{array}\right]\left[\begin{array}{cc}
\tilde{A}^{2} & \tilde{A}\tilde{C}\\
\tilde{A}\tilde{C} & \tilde{C}^{2}
\end{array}\right]\left[\begin{array}{c}
\tilde{\Phi}'\\
\tilde{\psi}'
\end{array}\right]\right)+p\tilde{\Phi}'\tilde{\psi}+2q\tilde{\Phi}\tilde{\psi}+m_{1}\tilde{\Phi}^{2}+m_{2}\tilde{\psi}^{2}\bigg\}\\
=\varepsilon^{2}\int\mathrm{d}^{4}x\,\frac{a^{2}}{2}\Big\{ & \tilde{u}\left(\mathbf{f}^{{\rm T}}\right)'\mathbf{K}\mathbf{f}'+p\tilde{\Phi}'\tilde{\psi}+2q\tilde{\Phi}\tilde{\psi}+m_{1}\tilde{\Phi}^{2}+m_{2}\tilde{\psi}^{2}\Big\},
\end{align}
where we have defined
\begin{equation}
\mathbf{f}=\left[\begin{array}{c}
\tilde{\Phi}\\
\tilde{\psi}
\end{array}\right],\quad\mathbf{K}=\left[\begin{array}{cc}
\tilde{A}^{2} & \tilde{A}\tilde{C}\\
\tilde{A}\tilde{C} & \tilde{C}^{2}
\end{array}\right].
\end{equation}
We can diagonalize the kinetic matrix as:
\begin{equation}
\mathbf{K}=\left[\begin{array}{cc}
\tilde{A} & -\tilde{C}\\
\tilde{C} & \tilde{A}
\end{array}\right]\left[\begin{array}{cc}
1 & 0\\
0 & 0
\end{array}\right]\left[\begin{array}{cc}
\tilde{A} & \tilde{C}\\
-\tilde{C} & \tilde{A}
\end{array}\right].
\end{equation}
So, if we define
\begin{equation}
\mathbf{Q}=\left[\begin{array}{cc}
\tilde{A} & \tilde{C}\\
-\tilde{C} & \tilde{A}
\end{array}\right],
\end{equation}
this suggests the field redefinition:
\begin{equation}
\tilde{\mathbf{f}}=\left[\begin{array}{c}
\zeta\\
\xi
\end{array}\right]=\mathbf{Q}\mathbf{f}=\left[\begin{array}{cc}
\tilde{A} & \tilde{C}\\
-\tilde{C} & \tilde{A}
\end{array}\right]\left[\begin{array}{c}
\tilde{\Phi}\\
\tilde{\psi}
\end{array}\right]=\left[\begin{array}{c}
\tilde{A}\tilde{\Phi}+\tilde{C}\tilde{\psi}\\
-\tilde{C}\tilde{\Phi}+\tilde{A}\tilde{\psi}
\end{array}\right].
\end{equation}
Inserting this into $S^{\left(2\right)}$ diagonalizes the kinetic
term. After simplifying and integrating by parts, we can write it
as:
\begin{align}
S^{\left(2\right)} & =\varepsilon^{2}\int\mathrm{d}^{4}x\,\frac{a^{2}}{2}\bigg\{\tilde{u}\left(\tilde{\mathbf{f}}'\right)^{\mathrm{T}}\left[\begin{array}{cc}
1 & 0\\
0 & 0
\end{array}\right]\tilde{\mathbf{f}}'+\tilde{\mathbf{f}}^{\mathrm{T}}\mathbf{M}\tilde{\mathbf{f}}+\left(\tilde{\mathbf{f}}'\right)^{\mathrm{T}}\mathbf{N}\tilde{\mathbf{f}}\bigg\}\\
 & =\varepsilon^{2}\int\mathrm{d}^{4}x\,\frac{a^{2}}{2}\Big\{\tilde{u}\zeta'^{2}+\tilde{\mathbf{f}}^{\mathrm{T}}\mathbf{M}\tilde{\mathbf{f}}+\left(\tilde{\mathbf{f}}'\right)^{\mathrm{T}}\mathbf{N}\tilde{\mathbf{f}}\Big\},\label{eq:E.S2ftilde}
\end{align}
where, if we define for convenience
\begin{equation}
\mathcal{A}=\frac{1}{\left(\tilde{A}^{2}+\tilde{C}^{2}\right)},
\end{equation}
then
\begin{equation}
\mathbf{N}=\left[\begin{array}{cc}
N_{1} & N_{2}\\
N_{3} & N_{4}
\end{array}\right],
\end{equation}
with
\begin{align}
N_{1}=\: & \mathcal{A}^{2}\left[p\tilde{A}\tilde{C}-2\tilde{u}\frac{1}{\mathcal{A}}\left(\tilde{A}\tilde{A}'+\tilde{C}\tilde{C}'\right)\right],\\
N_{2}=\: & \mathcal{A}^{2}\left[p\tilde{A}^{2}-2\tilde{u}\frac{1}{\mathcal{A}}\left(\tilde{A}\tilde{C}'-\tilde{C}\tilde{A}'\right)\right],\\
N_{3}=\: & \mathcal{A}^{2}\left[-p\tilde{C}^{2}\right],\\
N_{4}=\: & \mathcal{A}^{2}\left[-p\tilde{A}\tilde{C}\right],
\end{align}
and
\begin{equation}
\mathbf{M}=\left[\begin{array}{cc}
M_{1} & M_{2}\\
M_{3} & M_{4}
\end{array}\right],
\end{equation}
with
\begin{align}
M_{1}=\: & \mathcal{A}^{2}\bigg[\tilde{u}\left(\tilde{A}\tilde{A}'+\tilde{C}\tilde{C}'\right)^{2}+m_{1}\tilde{A}^{2}+m_{2}\tilde{C}^{2}+2q\tilde{A}\tilde{C}+\frac{p}{\mathcal{A}}\tilde{C}\left(\mathcal{A}\tilde{A}\right)^{'}\bigg],\\
M_{2}=\: & \mathcal{A}^{2}\bigg[\left(\tilde{A}^{2}-\tilde{C}^{2}\right)\left(\tilde{u}\tilde{A}'\tilde{C}'+q\right)+\tilde{A}\tilde{C}\left(\tilde{u}\left[\tilde{C}'^{2}-\tilde{A}'^{2}\right]+m_{2}-m_{1}\right)+\frac{p}{\mathcal{A}}\tilde{A}\left(\mathcal{A}\tilde{A}\right)'\bigg],\\
M_{3}=\: & \mathcal{A}^{2}\bigg[\left(\tilde{A}^{2}-\tilde{C}^{2}\right)\left(\tilde{u}\tilde{A}'\tilde{C}'+q\right)+\tilde{A}\tilde{C}\left(\tilde{u}\left[\tilde{C}'^{2}-\tilde{A}'^{2}\right]+m_{2}-m_{1}\right)-\frac{p}{\mathcal{A}}\tilde{C}\left(\mathcal{A}\tilde{C}\right)'\bigg],\\
M_{4}=\: & \mathcal{A}^{2}\left[\tilde{u}\left(\tilde{A}\tilde{C}'-\tilde{C}\tilde{A}'\right)^{2}+m_{1}\tilde{C}^{2}+m_{2}\tilde{A}^{2}-2q\tilde{A}\tilde{C}-\frac{p}{\mathcal{A}}\tilde{A}\left(\mathcal{A}\tilde{C}\right)'\right].
\end{align}
So, if we further define
\begin{align}
c_{1}=\: & M_{1}-\mathcal{H}N_{1}-\frac{1}{2}N_{1}',\\
c_{2}=\: & M_{4}-\mathcal{H}N_{4}-\frac{1}{2}N_{4}',\\
c_{3}=\: & M_{2}+M_{3}-2\mathcal{H}N_{3}-N_{3}',\\
c_{4}=\: & N_{2}-N_{3},
\end{align}
then (\ref{eq:E.S2ftilde}) yields precisely (\ref{3.eq:S2zx}).

\section{Deep Antigravity Full Solution\label{a:Solution}}

Upon substituting the background solution (\ref{eq:4.p0sol}) for
$\phi_{0}$ from \cite{Bars2012a}, using the properties of Jacobi
elliptic functions and dropping the $\mathcal{O}\left(1\right)$ terms
in $\phi_{0}$, our full ODE (\ref{eq:4.PEOM}) for the perturbation
$\psi$ becomes:
\begin{equation}
0=\frac{{\rm d}^{2}}{{\rm d}\eta^{2}}\psi+\left[-4\frac{{\rm cn}\left(\eta,m\right)}{{\rm sn}\left(\eta,m\right){\rm dn}\left(\eta,m\right)}\right]\frac{{\rm d}}{{\rm d}\eta}\psi+\left[2\sqrt{2}\frac{{\rm cn}\left(\eta,m\right)}{{\rm sn}\left(\eta,m\right){\rm dn}\left(\eta,m\right)}\right]^{2}\psi.
\end{equation}
This has an exact solution:
\begin{align}
\psi= & \left(1-m^{2}{\rm sn}^{2}\left(\eta,m\right)\right)^{-\frac{13+{\rm i}\sqrt{23}}{2(1+{\rm i}\sqrt{23})}}\left(-{\rm sn}^{2}\left(\eta,m\right)\right)^{\frac{5}{4}+{\rm i}\frac{\sqrt{7}}{4}}\nonumber \\
 & \times\bigg\{\mathscr{C}_{1}{\rm H}\bigg(1-\frac{1}{m^{2}},\frac{[2+{\rm i}(\sqrt{7}+\sqrt{23})][4m^{2}-2+{\rm i}(\sqrt{23}-\sqrt{7})]}{32m^{2}},\nonumber \\
 & \quad\quad1+{\rm i}\frac{\sqrt{7}+\sqrt{23}}{4},\frac{1}{2}+{\rm i}\frac{\sqrt{7}+\sqrt{23}}{4},\frac{1}{2},1+{\rm i}\frac{\sqrt{7}}{2},1-{\rm sn}^{2}\left(\eta,m\right)\bigg)\nonumber \\
 & +\mathscr{C}_{2}{\rm cn}\left(\eta,m\right){\rm H}\bigg(1-\frac{1}{m^{2}},[10+3{\rm i}(\sqrt{7}+\sqrt{23})]\nonumber \\
 & \quad\quad\times\frac{[1324m^{2}-477-9\sqrt{7}\sqrt{23}+{\rm i}(162\sqrt{23}-192{\rm i}\sqrt{7})]}{10592m^{2}},\nonumber \\
 & \quad\quad\frac{3}{2}+{\rm i}\frac{\sqrt{7}+\sqrt{23}}{4},1+{\rm i}\frac{\sqrt{7}+\sqrt{23}}{4},\frac{3}{2},1+{\rm i}\frac{\sqrt{7}}{2},1-{\rm sn}^{2}\left(\eta,m\right)\bigg)\bigg\},
\end{align}
where ${\rm H}$ is the Heun general function, and $\mathscr{C}_{1}$
and $\mathscr{C}_{2}$ are constants. 

\bibliographystyle{apsrev}
\bibliography{references}

\end{document}